# Directed exfoliating and ordered stacking of transition-metal-dichalcogenides

Yanshuang Li[1,2], Xiuhua Xie[1*], Binghui Li[1], Xiaoli Sun[3*], Yichen Yang[4], Jishan Liu[4,5*], Jiying Feng[6], Ying Zhou[6], Yuanzheng Li[6], Weizhen Liu[6], Shuangpeng Wang[7], Wei Wang[7], Huan Zeng[1,2], Zhenzhong Zhang[8], Dawei Shen[4], Dezhen Shen[1*]

[1]State Key Laboratory of Luminescence and Applications, Changchun Institute of Optics, Fine Mechanics and Physics, Chinese Academy of Sciences, No. 3888 Dongnanhu Road, Changchun, 130033, People's Republic of China

[2]University of Chinese Academy of Sciences, Beijing 100049, People's Republic of China

[3]Institute of Theoretical Chemistry, Jilin University, Changchun 130023, People's Republic of China

[4]Center for Excellence in Superconducting Electronics, State Key Laboratory of Functional Materials for Informatics, Shanghai Institute of Microsystem and Information Technology, Chinese Academy of Sciences, Shanghai 200050, China

[5]Center of Materials Science and Optoelectronics Engineering, University of Chinese Academy of Sciences, Beijing 100049, China

[6]Key Laboratory of UV-Emitting Materials and Technology, Ministry of Education, Northeast Normal University, Changchun 130024, China

[7]MOE Joint Key Laboratory, Institute of Applied Physics and Materials Engineering and Department of Physics and Chemistry, Faculty of Science and Technology, University of Macau, Macao SAR 999078, P. R. China

[8]School of Microelectronics, Dalian University of Technology, Dalian, 116024 China

*Corresponding authors. Email: xiexh@ciomp.ac.cn; sunxiaoli@jlu.edu.cn; jishanliu@mail.sim.ac.cn; shendz@ciomp.ac.cn

**Two-dimensional van der Waals crystals arise limitless scope for designing novel combinations of physical properties via controlling the stacking order or twist angle of individual layers. Lattice orientation between stacked monolayers is significant not only for the engineering symmetry breaking but also for the study of many-body quantum phases and band topology. So far the state-of-art exfoliation approaches focusing on achievements of quality, size, yield, and scalability while lacking sufficient information on lattice orientation. Consequently, interlayer alignment is usually determined by later experiments, such as second harmonic generation spectroscopy, which increased the number of trials and errors for a designed artificial ordering and hampered the efficiency of systematic study. Herein, we report a lattice orientation distinguishable exfoliation method via gold favor epitaxy along the specific atomic step edges, meanwhile, fulfill requirements of high-quality, large-size, and high-yield of monolayers. Hexagonal- and rhombohedral-stacking configurations of bilayer transition metal dichalcogenides are built directly at once result of foreseeing the lattice orientation. Optical spectroscopy, electron diffraction, and angle-resolved photoemission spectroscopy are used to study crystals quality, symmetric breaking, and band tuning, which supports the exfoliating mechanism we proposed. This strategy shows the ability for facilitates the development of ordering stacking especially for multilayers assembling in the future.**

Disassembling and reassembling two-dimensional (2D) materials into expected lattice configurations, give rise to intriguing quantum phases. As a consequence of stacking two layers of van der Waals (vdWs) materials, whether a lattice mismatch or twist angle existed, there is an additional periodicity, moiré superlattices[1,2], which can alter the low-energy electronic structure and hence the particles correlations. Various correlated and topological phases were firstly observed in twisted bilayer graphene at the magic angle, such as, unconventional superconductivity[3], correlated insulators[4], orbital Chern ferromagnets[5], to name a few. As complementary to graphene-based moiré systems, transition metal dichalcogenides (TMDs), notably $MoS_2$, $WS_2$, $MoSe_2$, $WSe_2$, with homojunctions and heterojunctions, have more advantages for realizing model quantum Hamiltonians. For example, Hubbard model physics has been simulated in twist-TMDs, due to the simpler moiré bands structure[6-8]. Generalized Wigner crystal has also been explored[9-11], attributed to the moiré flat bands, which are generally considered to be formed in a larger range of twist-angle, tolerating for the twist-angle disorder[12]. Moreover, benefit from strong matter–light interactions and deep tunable moiré potentials (100 to 200 meV)[13], bosonic quantum many-body effects are formed at higher temperatures[14,15]. TMDs-based moiré systems, actually, are robust quantum simulators for the study of strong correlations and band topology[16-21].

Unlike twist bilayer graphene, there are two distinct stacking orientation in TMDs semiconductor bilayers, which are R-type stacking (also labeled AA stacking, 0° rotation between two monolayers) and H-type stacking (also labeled AB stacking, 60° or 180° rotation between two monolayers). In homobilayers, H-type bilayer stacking has inversion symmetry, while R-type stacking broken. As twisting between the layers occurred, three kinds of high-symmetry stacking orders arise from moiré superlattice have been put forward[13,22]. These orders emerge very different energy and topological nature in near R- and H-type stacking, which means twist-TMDs have an extra degree of freedom for controlling. Stacking orientation, basically, plays a significant role in other vdWs materials as well, for instance, emergent ferroelectric order in parallel stacked bilayer boron nitride (BN)[23-25] and engineering interface polarization based on in-plane inversion symmetry broken[26]. As of late, interfacial ferroelectricity has been found in R-stacked bilayer TMDs[27,28]. However, it is tricky to predetermine the stacking orientation during the stack-up process, as the absence distinguishing of characteristic edges (especially for metal or chalcogen atoms ended zigzag edge) via exfoliated-edge alignment and second harmonic generation (SHG) methods thus far.

Previous skillful efforts for exfoliating vdWs materials into monolayer mainly focusing on crystal quality, size, yield, and scalability. Examples include solution phase exfoliation[29], using cation as the intercalant, has scalable and processable advantages at the expense of sizes and quality. Different growth approaches, including chemical vapor deposition (CVD) and molecular beam epitaxy (MBE)[30,31], can produce monolayers in wafer scale, but polycrystalline and high defect density are still unsurpassed. The original and facile Scotch-tape approach has enabled the highest quality of the monolayers, yet limited by low yield and small size. Recently, utilizing stronger vdWs adhesion between metal and 2D materials to overcome the interlayer vdWs force, high-quality monolayers with large-area, near-unity yield have been produced[32-35]. Unfortunately, all of these exfoliation strategies essentially lack sufficient information on the lattice orientation of peeled monolayers. This inadequacy motivating us to develop innovative peeling approaches with identifiable lattice orientation simultaneously meets high-quality, large-size, high-yield, which will be expediting further advancement in the field of engineering of symmetry breaking including moiré physics.

As rethinking the process of peeling off, we perceived that disassembling vdWs bulk crystals into monolayers usually overcomes the weaker interlayer attraction in the out-of-plane direction, while in the in-plane it lacks the differentiation along the specific crystal orientation. For monolayer, described points group of TMDs of 2H-phase degraded to $D_{3h}$, lower symmetry compares to $D_{6h}$ of bulk. It contains a threefold $C_3$ rotational symmetry with a $S_3$ mirror-rotation symmetry, which means there are two typical orientations of armchair (AC) and zigzag (ZZ). Among ZZ orientation, there have two types of edge: metal-terminated edge and chalcogen-terminated edge. In principle, these three kinds of edges should have very different adsorption energies or bond enthalpy, which will provide the possibility to peel the vdWs crystals from a specific edge.

Here, we demonstrate an edge-locked mechanical exfoliation method for bulk TMDs with discernible orientation, high-quality, large-dimension, and high productivity. This method is practicable for producing a range of orientated monolayers vdWs materials, so long as their parent counterparts have the typical atomic edges. Using $MoS_2$ as a representative system, a pivotal start is illustrated in Fig. 1a, gold (Au) atoms preferred epitaxy along the Mo-terminated ZZ edge, owing to the higher bond strength compared to edges of S-terminated ZZ or AC (details can be seen in the theory section below). This kind of Au-locking edge will play the role of grasping during the process of peel-off, which inducing the monolayer tearing along AC orientation from Mo- to S-terminated of ZZ.

## Results

**Au-epitaxy aligns atomic-step edges and theoretical analysis.** In order to accurately control the flux of Au atoms reaching the cleaved TMDs surface and to have a sufficient surface relaxation process, we used MBE in this work (Fig. 1a). Owing to the low beam flux and ultrahigh-vacuum conditions (see "Methods"), it can be guaranteed no extra surface defects be introduced as usual evaporation methods (e.g. thermal, sputtering). Freshly cleaved bulk TMDs surface has flat terraces, which parted by atomic steps. These atomic steps, essentially, are one-dimensional (1D) defects, serving as a nucleation site, which tends to adsorb or bind heteroatoms to reduce the total energy of the system[36]. Therefore, during the epitaxy of Au crystals, adatoms are preferring bonding along these active 1D edges and then crystal growth. The surface morphology, after the initial growth of Au nanocrystalline on the surface of $MoS_2$, observed via scanning electron microscopy (SEM) and atomic force microscopy (AFM), supports the inference of the nucleation process. As can be seen in Fig. 1b, c, besides the sporadic random distribution on the entire surface, there are also obvious Au nanocrystal grains arranged in a specific direction. In the top view, Au nanocrystal grains appear equilateral triangular, which means that the crystal orientation is (111). Interestingly, these arranged triangular crystal grains are aligned along the side of each triangle and form an angle of 120° at the point of convergence, which is a consequence of the $C_3$ symmetry of $MoS_2$. This kind of alignment feature indicates that Au atoms nucleate and grow at the ZZ atomic step edges (see the theory part below. They are actually Mo-terminated ZZ edges). We, in fact, have examined the morphology of many surfaces after initial epitaxy and found that only in rare cases the triangular grains are arranged along the angle bisector of the triangle (Supplementary Fig. S1), which means that the Au atoms hardly grow along the AC edges. Atomic steps with arranging Au nanocrystal grains are further analysed by AFM (Fig. 1d, e). The line contour corresponding to the white arrow in Fig. 1e shows that the height difference at the steps is 0.65 nm, which equals the thickness of the $MoS_2$ monolayer. These Au bonded atomic steps are locked during the subsequent peeling process, inducing the oriented peeling.

In order to fully fathom the physical process of Au atom lateral epitaxy, we carried out first-principles density functional theory (DFT) calculations relating to the initial binding of Au atoms onto the atomic step edges (AC and two kinds ZZ) of $MoS_2$ and the subsequent alignment between Au nanocrystal grains and atomic edges.

The docking energy ($E_d$) for Au(111) on $MoS_2$ step contains two parts: the surface and the step contact. The surface energy ($E_s$) only contains the contribution that Au(111) adsorbed on the bottom $MoS_2$ surface and vdWs energy interaction is in the majority, while the step contact energy ($E_c$) infers the connection between the S or Mo atom of $MoS_2$ step and Au atom of Au(111) closely which is adjacent to $MoS_2$ stage. The step contact energy per length ($\varepsilon$) was defined $E_c/L$ to describe the ability of docking between $MoS_2$ and Au(111). L (Å) represent the length of edges, for ZZ steps, L = 22.75 Å, while for AC steps, L = 31.74 Å. Specific formulas for $\varepsilon$, $E_s$, $E_d$, and $E_c$ can be found in Supplementary Note 8. The lower $\varepsilon$ (eV/Å), greater ability of docking between $MoS_2$ and Au(111).

A comparison of the $\varepsilon$ values for different step structures of the $MoS_2$ bilayer is presented in Fig. 2d and Supplementary Table 1. Figure 2a-c shows the top and side views of the DFT-optimized structures of the contact configurations between Au edges and $MoS_2$ substrate. The results show that the $\varepsilon$ values for the ZZ-Mo direction with A and B steps (Supplementary Fig. S8a and S8d) are −1.11 and −0.67 eV/Å, respectively, while the corresponding values for the AC direction with C and D steps (Supplementary Fig. S8e and S8f) are −0.64 and −0.64 eV/Å, respectively. However, the $\varepsilon$ values for the ZZ-S direction with A and B steps (Supplementary Fig. S8b and S8c) are −0.19 and −0.07 eV/Å, respectively. It is clear that the contact of ZZ direction prefers A step, while AC presents the same $\varepsilon$ of the two steps ($\varepsilon$ = −0.64/Å). Moreover, the coupling between ZZ-Mo edge is energetically far greater than ZZ-S edge, which is also similar to the situation described by Yang et al.[37]. Combing the optimized structures AC direction and the energies above, it seems that the contact is mostly contributed by Mo, $MoS_2$ layer would be stripped from ZZ-Mo direction. To test the impact of the established model on $\varepsilon$, the model was expanded, i.e. the width was expanded of AC direction, it was found that $\varepsilon$ has almost no change in value (see Supplementary Fig. S8).

From Fig. 2, the MoS$_2$ step docked by triangular Au(111) presents two scenes: 1) the triangles across the steps, AC edges, i.e., one corner points along step edge, and one side of the triangles perpendicular to it; 2) One side of a triangle in contact with the step. i.e., one corner contacts the step edge, and one side of the triangles parallel to it; ZZ edge. For both the two scenes, arrangement of the directional triangle is ordered and oriented. That is, the orientation of Au(111) on MoS$_2$ can be used to characterize the edge structure (AC or ZZ) or stack MoS$_2$ directionally. These scenes present on the step edge have been coincided with the submonolayer MoS$_2$ growth on highly oriented pyrolytic graphite (HOPG)[38] and is in accordance with our SEM and AFM observations shown in Fig. 1 and Supplementary Fig. S1. Moreover, a comparison of the $\varepsilon$ values for different step structures of Au (111)-WS$_2$, Au(111)-MoSe$_2$ and Au(111)-WSe$_2$, as well as Au (111)-MoS$_2$ were given in Supplementary Table 1. The same conclusions are exhibited for MX$_2$ (M = Mo, W; X = S, Se), that is, step contact is mostly contributed by Mo or W. It indicates that MX$_2$ layers would be stripped from the ZZ-M direction.

**Oriented exfoliation and monolayers optical characterization.** The schematic diagram of the core procedures of mechanical exfoliation is illustrated in Fig. 3a, where the critical is the first procedure, namely, edge epitaxy of Au atoms in the initial growth stage. Relatively low beam equivalent pressure (BEP), $1.4 \times 10^{-8}$ Torr, makes the adsorbed Au atoms can fully relax, inducing the Mo-terminated ZZ edges bonded uniformly. Uneven bonding edges induced peeling results can be seen in Supplementary Fig. S5. Then, the BEP is slightly increased, and relatively quickly epitaxially forms a continuous Au film, which plays a role in the background vdWs adhesion with out-of-plane direction, while protecting the stripped layers from contamination, during the subsequent peeling process. The thermal release tape is gently pressed on top of the Au film to form even and close contact. Peel off the top layer of TMDs from one side, at this time, the monolayer is exfoliated along the tearing AC edges under the combined action of the adhesion forces from two directions, viz., out-of-plane (from the topmost Au film) and in-plane (from the Au bonded ZZ edges). After that, transferring it onto the target substrate, and removing the thermal release tape by heating to the right temperature. Using mild iodine/potassium iodide ($I_2$/KI) solution to etch away the Au film and dissolving possible residues of the thermal release tape with acetone. Finally, the large-area monolayer TMDs with straight AC edges are obtained. Its lateral dimensions reach the sub-millimeter scales (Fig. 3c), which are limited mainly by the sizes of the terraces separated by atomic steps. The peeled surface layers (the inset of Fig. 3d) can be completely transferred to the substrate by using this method. Since the second-order susceptibility tensor $\chi^{(2)}$ for non-centrosymmetric monolayer TMDs along the AC direction is non-vanishing, the linear polarization-dependent SHG can identify the orientation and symmetry[39]. As shown in Fig. 3d, the polar diagram of the SHG intensity represents the AC orientation in parallel with torn edges, useful for precise alignment between each layer in homo- and hetero-stacking. Substantially equal polarization SHG intensity distribution also illustrates that the strain that can cause the lattice vector change is insignificant after this peeling-transfer process. Notably, along the peeling direction, the number of layers generally presents the characteristics of the transition from monolayer to bilayer and even multilayer, which is the consequence of the 2H stacking order (the atomic edges of the adjacent terraces have different termination atoms), as pointed up in Fig. 3b. This is also served as orientation recognition for the following artificial stacking. The resulting optical images and SHG intensity polar diagrams of achieved monolayer of $WS_2$, $MoSe_2$, and $WSe_2$ are shown in Supplementary Fig. S2, which exhibit the same situation.

Photoluminescence (PL) and Raman spectroscopy is used to evaluate the crystal quality of TMDs monolayers. As shown in Fig. 4a, there are two distinct emission peaks at room temperature, located 1.86 eV and 2.02 eV, corresponding to A- and B-exciton respectively, for a monolayer $MoS_2$ on $Si/SiO_2$ (285 nm) substrate. Two typical Raman vibrational modes are observed in Fig. 4b, in which the spacing between $E^1_{2g}$ and $A_{1g}$ is 19.8 cm$^{-1}$ (determined by interlayer interactions), supporting one layer thickness. PL and Raman 2D imaging of the monolayer $MoS_2$ is shown in Supplementary Fig. S4, in which a spatial homogeneous of PL peak and phonon wavenumber spacing are displayed. Results for $WS_2$, $MoSe_2$, and $WSe_2$ can be found in Supplementary Fig. S3, confirming the monolayers with high-quality.

**Artificial stacking and energy band tuning.** These stripped monolayers provide the prerequisite for stacking orientation known in advance. Based on the judgment of orientation, mentioned before, two types of bilayer stacking ordering (H-type and R-type) are intentionally constructed, shown in Fig. 5. The detailed stacking process is shown in Supplementary Fig. S7, of which the upper monolayers are transferred on transparent Polydimethylsiloxane (PDMS) directly for a stamp dry transfer method. H-type stacking homobilayer $MoS_2$ exhibits inversion symmetry, resulting in a weaker SHG intensity compared to the monolayer region, as shown in Fig. 5a. Meanwhile, R-type stacking has the opposite situation, inversion symmetry breaking homobilayer has stronger SHG intensity, due to interference coherent enhanced from the individual monolayers (Fig. 5b). Pre-identification of crystal orientation improves the efficiency of symmetry control in our case.

Furthermore, straight and longer armchair edges, obtained in our method, provide a more accurate alignment basis for twisted stacking. As illustrated in the inset of Fig. 6, a twisted bilayer $MoS_2$ with the designed angle of 3.03° is fabricated by angular rotating between two AC edges. The actual twist-angle, 3.39° (Fig. 6), is confirmed by selected-area electron diffraction (SAED), which is close to the designed rotation angle, indicating the accuracy by using longer AC edge. Additionally, surrounding each pair of hexagonal diffraction spots, high order satellite spots have been observed, which caused by interlayer interaction and atomic reconstruction of twisted-bilayer.

The large size high quality stacked bilayer allows us to obtain the electronic structure with standard angle-resolved photoemission spectroscopy (ARPES) technique using a conventional setup with a hemispherical electron energy analyzer without microscopic capabilities. Figure 7 presents M-Γ-K slices, showing the important features within 4 eV under the Fermi level for the H-type (Fig. 7a, c, e) and R-type (Fig. 7b, d, f) stacked $MoS_2$ bilayers. All features of the valance bands are well resolved, implying the method we developed could well keep the sample surface from deteriorations and contaminations. The band structure over the entire surface-Brillouin zone reveals that the valence band maximum (VBM) at Γ is indeed the highest occupied state for both the H-type and R-type stacked bilayers, affirming the indirect nature of the bilayer $MoS_2$ bandgap. Note that our ARPES data clearly show an energy difference between the valence band maximum at Γ and K for H- and R-stacked bilayers, as shown in Fig. 7a and b, and their corresponding Second-derivative spectra Fig. 7c and d. The energy difference is further determined in the corresponding energy distribution curves (EDCs), as shown in Fig. 7e and f. We know that the VBM at Γ is derived from out-of-plane S $p_z$ and Mo $d_z^2$ orbitals, while the VBM at K is mostly derived from Mo $d_{x^2-y^2/xy}$ orbitals[40]. Thus, the energy of the valence band maxima at Γ is sensitive to the out-of-plane interlayer coupling, as compared to the K point. In our case, according to previous reports[41,42], the R-stacked bilayer has a smaller interlayer spacing compared to the H-stacked one, which inducing stronger interlayer coupling between the out-of-plane orbitals in R-type stacking order, and giving rise to an upshift in the energy of the Γ point, as observed in Fig. 7.

In summary, we develop an orientation recognizable mechanical exfoliation method for disassembly of 2D vdWs materials. The obtained monolayers have high-quality, large-dimension, and are torn along the armchair orientation, which is verified by PL, Raman, and SHG. As be known to the orientation information in advance, the artificial stacking orderings (R-type and H-type) with different symmetry and band structures have been successfully realized without accidents, which demonstrated by SHG mapping and ARPES. This strategy is not only effective for $MoS_2$, $WS_2$, $MoSe_2$, $WSe_2$, confirmed by experiments, but also promising for other TMDs, such as Nb-, Ta-, or Ti-based TMDs, in which as long as choosing the appropriate metals meet the requirements of lateral epitaxy and damage-free etching. The clear-cut orientation of monolayers will make the research for twistronics, including moire physics or band topology in vdWs stacking, more facilitated.

**Methods**

**Gold atoms epitaxy on the surface of transition metal dichalcogenides (TMDs) bulk crystals.** Freshly cleaved TMDs bulk crystals (6 Carbon Technology, Shenzhen) were transferred on the cleaned silicon (Si) wafer surface by using thermal release tape (TRT) with releasing temperature of 90 °C. Previous to the gold (Au) atoms epitaxy in the main chamber of molecular beam epitaxy (MBE) system (P600, DCA Instruments, Finland), samples were transported to an ultrahigh-vacuum (UHV) subsystem (base pressure of $3 \times 10^{-9}$ Torr), vacuum interconnected with the main chamber, for surface cleaning under the condition of heating at 350°C for 12 hours. During the epitaxy, the K-cell contains Au (6N grade) was kept at 1260 °C to form a steady Au atom beam equivalent pressure (BEP), which is $1.4 \times 10^{-8}$ Torr, while the TMDs/Si substrate was kept at 200 °C. The temperature of the K-cell was increasing to 1280 °C for afterward Au film growth.

**Mechanical exfoliation.** Firstly, TRT (releasing temperature 90 °C) gently pressed onto the surface of prepared Au/TMDs/Si following a peeling-off process. Under the action of the combination of adhesion from two directions, topmost layers of bulk TMDs including oriented monolayers were stripped, which attached to the Au surface. The TRT/Au/TMDs subsequently transferred onto the desired substrate. Secondly, the TRT was removed by heating at 90 °C. The Au film was dissolved by Potassium Iodide (KI) and Iodide ($I_2$) solution (35 g KI and 13 g $I_2$ dissolved in 100 ml deionized water). Finally, the monolayer of TMDs was further rinsed with deionized water and acetone for 5 minutes and dried by $N_2$ flow.

**Morphological and crystal structure characterization.** The surface topography of bulk $MoS_2$, at the initial stage of Au epitaxy, was measured by atomic force microscope (Multimode8, Bruck Instruments, Germany) and scanning electron microscope (S-4800, Hitachi Instruments, Japan). The selected area electron diffraction (SAED) of $MoS_2$ bilayer was measured by Field Emission Transmission

Electron Microscope (TEM) (Tecnai G2 F20, FEI, American) with 200 kV accelerating voltage.

**Optical characterization.** The optical images were obtained in a custom-built microscopy system by using a 20× objective lens. Photoluminescence (PL), Raman spectra and polarization-dependent second harmonic generation (SHG) performed at room temperature on a Laser Scanning System (ScanPro Advance, Metatest, China). The laser-focused spot is about 1 um diameter by a 100× objective lens (Olympus Corporation, Japan). The wavelength of the excitation source for PL and Raman was 532.18 nm with power of 0.1 mW. Polarization-dependent SHG was performed using a 1064 nm picosecond laser (10 ps), with two polarizers (polarizer and analyzer) equipped with a rotating half-wave plat between them.

**SAED sample preparation.** Mechanical exfoliation $MoS_2$ monolayers were firstly prepared onto polydimethylsiloxane (PDMS). 50 µL polycarbonate (PC) solution (8% weight in trichloromethane) dropped on a glass slide and self-leveling a PC film. Two of the $MoS_2$ monolayers were stacked from PDMS to the PC film by heating to 80 °C for 10 minutes. The twisted angle is about 3.03° by aligning the straight torn edges. Then the PC film was peeled from the glass slide and placed onto PDMS forming a stamp of twist-bilayer $MoS_2$/PC/PDMS. Finally, the twist-bilayer $MoS_2$ was transferred onto a carbon-supported copper grid by heating to 180 °C for 2 minutes. The PC was removed by trichloromethane and acetone.

**SHG sample preparation.** Two groups of mechanical exfoliation $MoS_2$ monolayers were firstly prepared onto PDMS and $Si/SiO_2$ (285 nm), respectively. Then aligned stacking them from PDMS to $Si/SiO_2$ (285 nm) with a rotation angle close to 0° or 180° by heating 80 °C for 10 minutes. Stacking ordering was predetermined as showing in Supplementary Fig. S7.

**Angle-resolved photoemission spectroscopy measurement.** High-resolution ARPES measurements were performed with incident photon energies of 140 eV at beamline 03U of Shanghai synchrotron radiation facility (SSRF). The ARPES system is equipped with a Scienta-Omicron DA30 electron energy analyzer. During measurement, the vacuum was better than $8.0 \times 10^{-11}$ Torr and temperature was kept at around 15 K. Energy resolutions were better than 20 meV. The beam size of the synchrotron radiation is roughly under 100 micrometers. For ARPES measurement, the samples were transferred onto an Au/MgO substrate. An annealing process lower than 300 °C for 12 h was performed to clean the surfaces after loading them into a high vacuum chamber.

**Density functional theory (DFT) calculations**

**Calculation Methods.** DFT calculations were consulted by the plane-wave method as implemented in the Vienna Ab Initio Simulation Package (VASP)[43,44]. The electron exchange-correlation and the ion-electron interactions were treated by Perdew−Burke−Ernzerhof functionals (PBE)[45] and projector-augmented wave (PAW)[46] potentials, and the van der Waals (vdW) corrections were considered by zero damping DFT-D3 method[47,48]. The energy cutoff of 400 eV for the plane-waves was employed, and the convergence criteria energy and force were set to $10^{-4}$ eV and 0.1 eV/Å for structural relaxation. Only the Gamma point was used due to large supercell. The periodic lattice perpendicular to the plane direction was larger than 20 Å to avoid artificial interactions.

**Model establishment.** The optimized lattice constant of Au and $MoS_2$ are 4.16 and 3.18 Å respectively, which is in good agreement with the experimental values of 4.072 and 3.169 Å[49,50]. The Au (111)-$MoS_2$ overlayer were modeled by a $(8 \times 6\sqrt{3})$ Au(111) on $(7 \times 6\sqrt{3})$ $MoS_2$ bilayer to simulate the contact configurations between Au edges and $MoS_2$ substrate with a lattice mismatch of ~5.6%. $MoS_2$ bilayer with 168 molybdenum (Mo), 336 sulfur (S) atoms and top Au monolayer with 96 atoms as the initial structure (lattice parameters a = 22.85 Å, b = 31.74 Å). Then the overlayer was cut and docked to two kinds of step edges: armchair (AC) direction and Zigzag direction (ZZ-Mo and ZZ-S). 2 (named A step) or 3 (named B step) rows of MoS$_2$ along the ZZ direction and 3 (named C step) or 2.5 (named D step) rows along AC direction are remaining to simulate the step structure of MoS$_2$, followed by placing the remaining 5 and 3 rows Au of Au (111) after cutting onto ZZ and AC edges with geometric relaxation respectively. The stage is wide enough to reduce the fluctuation of step atoms on the MoS$_2$ surface. The ZZ edge of MoS$_2$ is parallel to the <110> direction of Au (111) and the AC edge of MoS$_2$ is parallel to the <112> direction of Au (111). The geometries have been presented in Fig. 2 and Supplementary Fig. S8. Moreover, we perform additional test calculations to further verify the reliability of the calculation model (Supplementary Fig. S8).

**Acknowledgments**

This research is mainly supported by the National Natural Science Foundation of China (Grant Nos. 11727902 and 62074146). S.P. and X.H. gratefully acknowledge support from Multi-Year Research Grants (MYRG2020-00207-IAPME) from Research & Development Office at University of Macau. X.L. acknowledges generous financial support by the National Natural Science Foundation of China (Grant Nos. 21903036). Y.Z. acknowledges support from Program of National Natural Science Foundation of China (Grant Nos.12004069). W.Z. acknowledges support from the Program of National Natural Science Foundation of China (Grant Nos. 12074060 and 11874104), the Fund from Jilin Province (no. YDZJ202101ZYTS133 and JJKH20211273KJ) and the Fundamental Research Funds for the Central Universities (no. 2412019BJ006 and 2412021ZD012). J.S. and D.W. acknowledge the fund suppot from National Natural Science Foundation of China (U2032208). Part of this research also used beam line 02B of the Shanghai Synchrotron Radiation Facility, which is supported by the ME2 project under Contract No. 11227902 from the National Natural Science Foundation of China.

**Author contributions**

X.H. conceived the project. X.H. and D.Z. supervised this work. X.H., Y.S. and B.H. designed the experiments. X.L. performed the DFT calculations. Y.S. and H. prepared the mechanical exfoliation samples. Y.S., J.Y., Y., and Y.Z. performed the Raman, PL and SHG measurements. X.H. and B.H. performed the gold growth by MBE. Y.S. and W. performed the AFM measurements. J.S. and Y.C. performed the ARPES measurement. X.H., Y.S., and H. prepared the stacking samples. B.H., Y.S., H., and Z.Z. performed the SEM and SAED measurements. X.H., Y.S., X.L., and J.S. wrote the manuscript with inputs from all authors. All authors discussed the results and commented on the paper.

**Competing interests**

The authors declare no competing interests.

**Data Availability**

All data that support the findings of this study within this paper and the Supplemental Information are available from the corresponding authors upon reasonable request.

# Figures

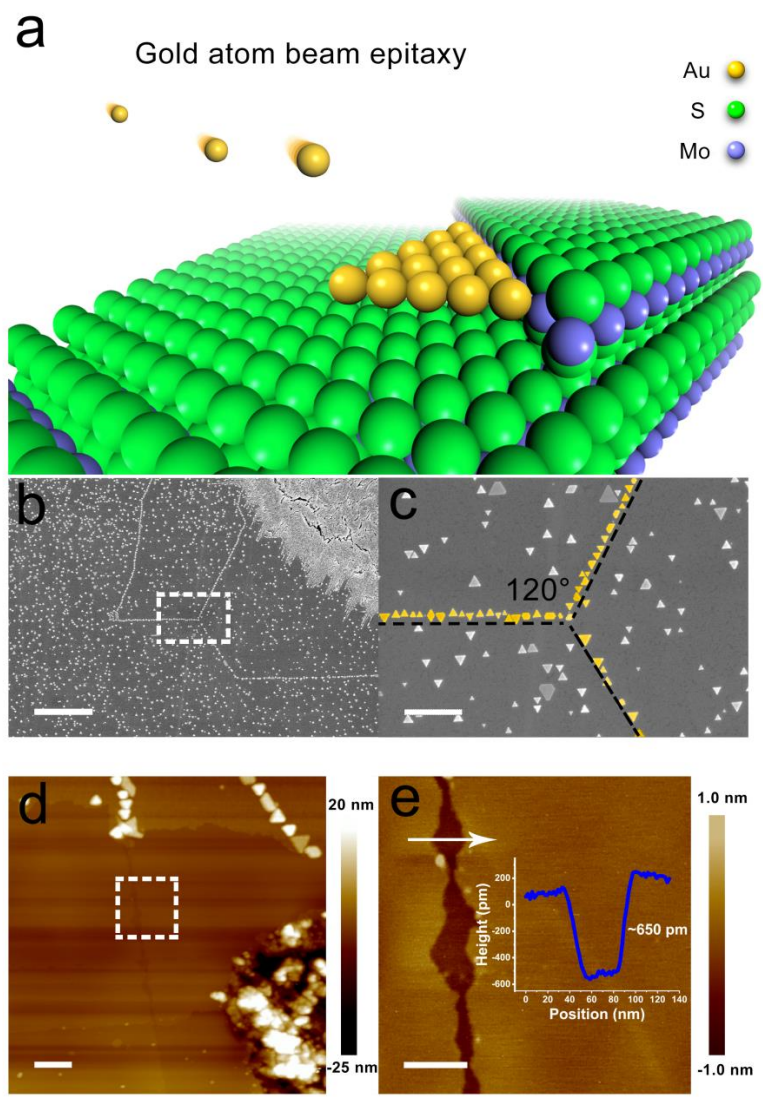

**Figure 1 | Gold epitaxy along the atomic steps on the surface of MoS$_2$ bulk crystals. a**, Schematic illustration of epitaxial growth of a triangle gold crystal along the zigzag atomic step of MoS$_2$ bulk crystals by MBE. **b**, **c**, False-colour Scanning Electron Microscopy images of the surface of MoS$_2$ bulk crystals. The yellow triangles (gold nanocrystals) aligned along with one side of the triangles forming Y-shaped lines with an angle of 120°. Scale bars, (**b**) 2 μm, (**c**) 400 nm. **d**, **e**, Atomic Force Microscope image is taken near the atomic step of MoS$_2$ surface. The height of the step as be shown in the blue curve, which is about 650 pm, indicates that it is the monolayer step of MoS$_2$. Scale bars, (**d**) 400 nm, (**e**) 100 nm.

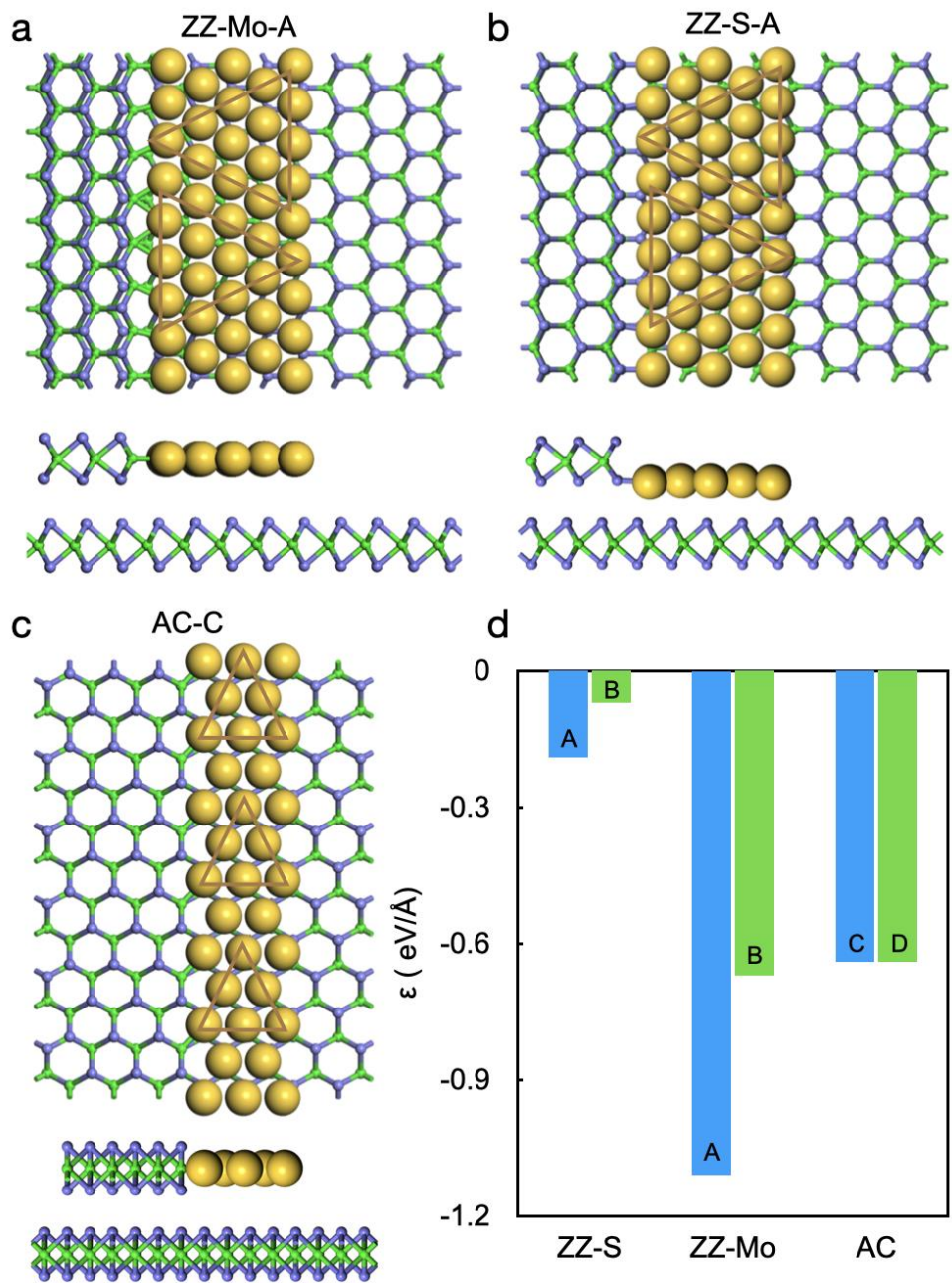

**Figure 2** | The ⟨110⟩ direction on Au(111) surface along two typical MoS$_2$ step edges and the ⟨112⟩ direction on Au(111) surface along one typical MoS$_2$ step edge. Atoms in golden, green and blue colors represent the Au, Mo and S atoms, respectively. **a**, **b**, coincide with Zigzag direction (ZZ-Mo and ZZ-S) with A step; **c**, armchair (AC) direction with C step. **d**, step contact energies per unit of the length of edges between the three edges of MoS$_2$ step and Au(111).

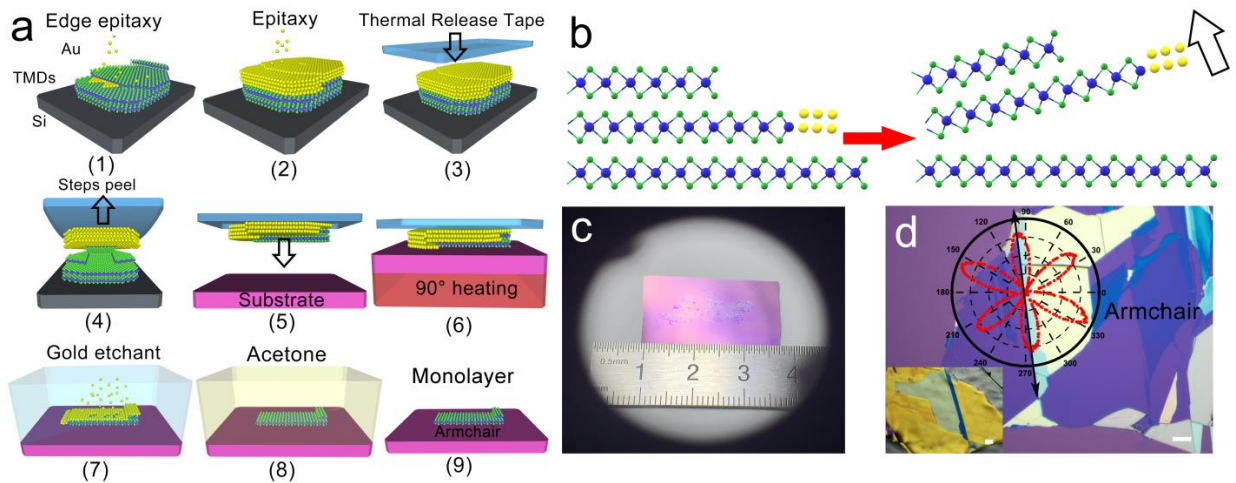

**Figure 3 | Methods of mechanical exfoliation along armchair direcion. a**, Schematic illustration: (1) Slowly epitaxy growth of gold on $MoS_2$ bulk crystals on silicon by MBE; (2) Epitaxy gold until gold atoms cover the surface; (3) Sticking thermal release tape to the surface of the gold film; (4) Pulling up thermal release tape and $MoS_2$ bulk crystals peeling off along the steps of the zigzag direction; (5) Sticking thermal release tape, gold and $MoS_2$ to the surface of the substrate; (6) 90° heating; (7) Etching gold by $KI/I_2$ solutions; (8) Cleaning residues by acetone; (9) Obtaining $MoS_2$ monolayers on the surface of $SiO_2$, which have a long armchair edge. **b**, Schematic illustration of mechanical exfoliation from zigzag-Mo to zigzag-S directions. **c**, **d**, Optical images of large areas and along the armchair direction of monolayer $MoS_2$ crystals. Scale bars, 50 μm.

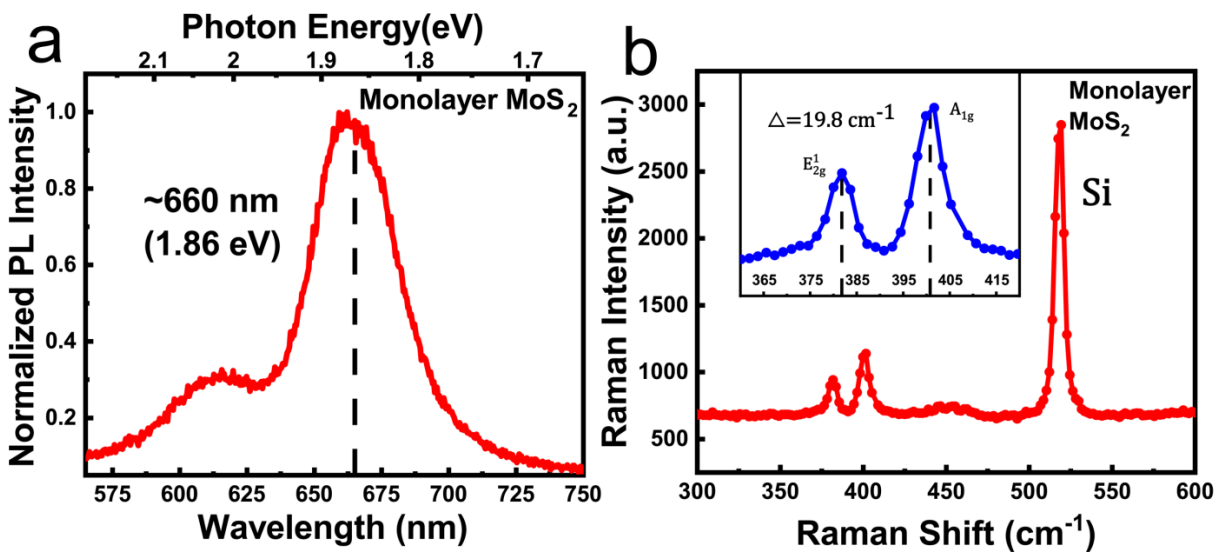

**Figure 4 | PL and Raman spectra of monolayer MoS$_2$ at room temperature with excitation laser wavelength is 532.18 nm. a**, Normalized PL spectra intensity of monolayer MoS$_2$. **b**, Raman shift of in-plane E$^1_{2g}$ mode and out-of-plane A$_{1g}$ mode of monolayer MoS$_2$.

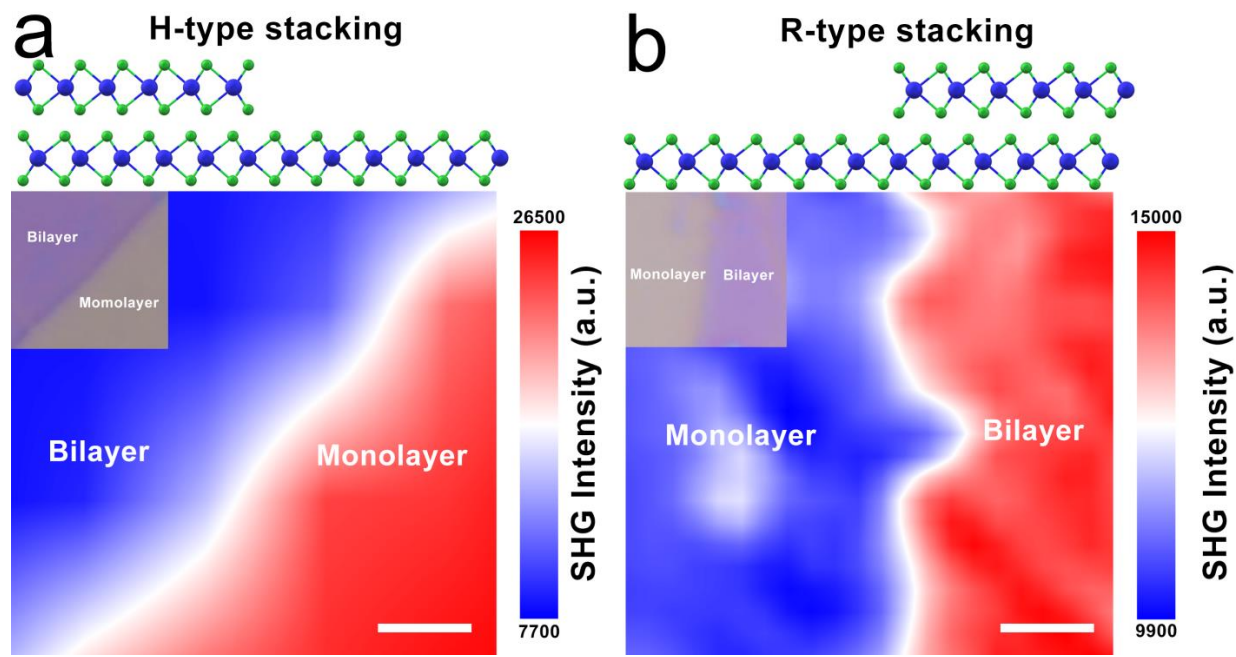

**Figure 5 | SHG mapping of H-type (anti-parallel) and R-type (parallel) stacking MoS$_2$ crystals. a**, Intensity of anti-parallel bilayer MoS$_2$ crystals is weaker than monolayer. **b**, Intensity of parallel bilayer MoS$_2$ crystals is stronger than monolayer. Scale bars, 1 μm.

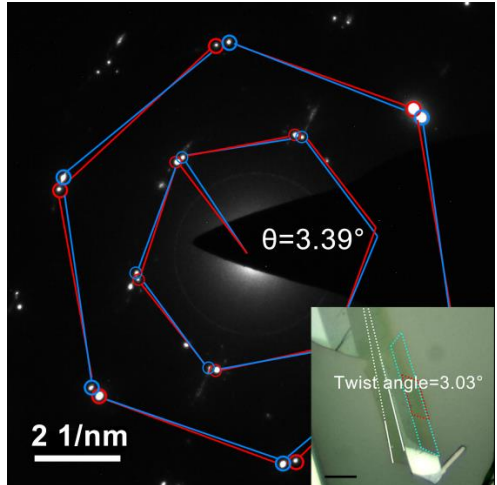

**Figure 6 | Selected area electron diffraction (SAED) image of twist bilayer MoS$_2$/MoS$_2$.** Twist angle is 3.39° in SAED image of bilayer MoS$_2$/MoS$_2$. And twist angle is 3.03° via measuring the angle between two straight edges in optical images of twist bilayer MoS$_2$/MoS$_2$ in the insert figure ( Scale bar, 10 μm).

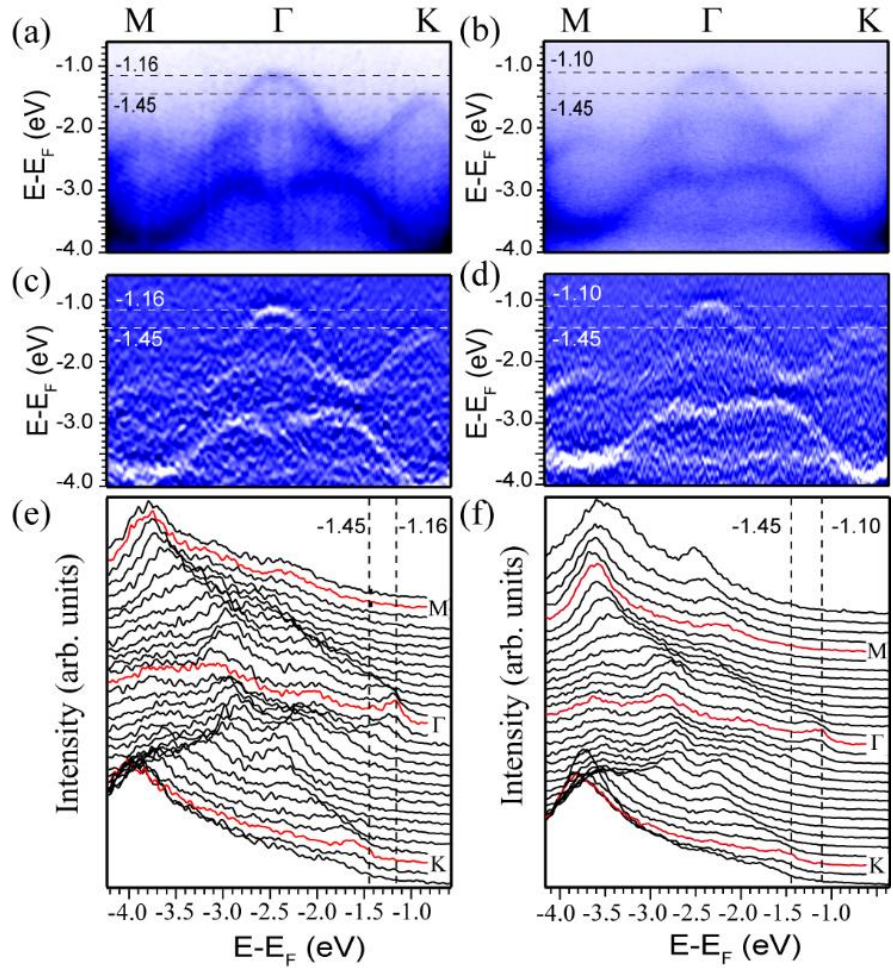

**Figure 7 | ARPES measurement of the sub-millimeter scale stacked bilayer MoS₂. a**, **b**, The band structure along the M-Γ-K high symmetry direction for H-type and R-type stacked $MoS_2$ bilayer. **c**, **d**, Second-derivative spectra of (**a**) and (**b**), respectively. **e**, **f**, The corresponding energy distribution curves (EDCs) along the high symmetry direction. The dashed lines indicate the energy difference between valence band maximum (VBM) at Γ and K for the H-type and R-type stacked $MoS_2$ bilayer.